\documentclass[fleqn,usenatbib]{mnras}

\usepackage{array}
\usepackage{newtxtext,newtxmath}
\usepackage{booktabs}
\usepackage{adjustbox}
\usepackage{ulem}
\usepackage{comment}
\usepackage{verbatim}

\usepackage[T1]{fontenc}
\usepackage{subfigure}
\newcommand{\ergcm}[1]{$\times 10^{#1}$ erg cm$^{-2}$ s$^{-1}$}
\DeclareRobustCommand{\VAN}[3]{#2}
\let\VANthebibliography\thebibliography
\def\thebibliography{\DeclareRobustCommand{\VAN}[3]{##3}\VANthebibliography}

\usepackage[switch, mathlines]{lineno}  
\usepackage{siunitx}
\usepackage{lipsum}
\usepackage{amsmath}

\usepackage{graphicx}	
\usepackage{amsmath}	
\usepackage{orcidlink}

\title[Neutrino and $\gamma$-ray emission from PKS~1725+123]{Zooming in on the GeV $\gamma$-ray flare of the blazar PKS 1725+123 with a multimessenger lens}

\author[SUVAS et al.]{
Suvas Chandra Chaudhary,\orcidlink{0000-0002-9126-1817}$^{1}$\thanks{2029666035@ufs4life.ac.za}
Saikat Das,\orcidlink{0000-0001-5796-225X}$^{2}$\thanks{saikatdas@ufl.edu}
Raj Prince,\orcidlink{0000-0002-1173-7310}$^{3}$\thanks{priraj@bhu.ac.in}
and Brian van Soelen \orcidlink{0000-0003-1873-7855} $^{1}$\thanks{VanSoelenB@ufs.ac.za}
\\  
$^{1}$Department of Physics, University of the Free State, 205 Nelson Mandela Dr., Bloemfontein, 9300, South Africa.\\
$^{2}$ Department of Physics, University of Florida, Gainesville, FL 32611, USA.\\
$^{3}$ Institute of Science, Banaras Hindu University (BHU), Varanasi-221005, India.\\ 
}

\pubyear{\the\year{}}

\usepackage{soul}
\usepackage[dvipsnames, svgnames]{xcolor}
\soulregister\citep7
\soulregister\citet7
\soulregister\cite7

\soulregister\ref7
\soulregister\eqref7
\soulregister\pageref7

\soulregister\autoref7      
\soulregister\cref7         
\soulregister\Cref7

\sethlcolor{yellow!50}

\begin{document}
\label{firstpage}
\pagerange{\pageref{firstpage}--\pageref{lastpage}}
\maketitle


\begin{abstract}
Blazars are promising sources of extragalactic high-energy astrophysical neutrinos, detected at energies $\gtrsim 10$ TeV by the IceCube neutrino observatory. Here, we report the first-ever broadband timing and spectral study of the flat-spectrum radio quasar PKS 1725+123, which has recently emerged as a compelling multimessenger target following its spatial association with the IceCube event IC-201021A. 
This triggered extensive follow-up observations from radio to VHE $\gamma$-rays, and a multi-episode flare was identified at a later time. During this period, the source exhibited high flux variability across all wavelengths. The {\it Fermi}-LAT analysis suggests rapid variability on timescales of less than 6 hours, implying a compact emission region with a radius of $\sim10^{16}$ cm. Our one-zone leptohadronic model shows that the high-energy $\gamma$-ray flux is produced by a combination of inverse-Compton scattering of external photons from the hot accretion disk and the broad-line region, while the X-ray emission is dominated by synchrotron self-Compton radiation from relativistic electrons. The secondary radiation from the hadronic cascade is found to be sub-dominant in the $\gamma$-ray regime, and the X-ray data constrain the maximum proton energy to $\sim 20$ PeV in the observer frame. Photopion production occurs predominantly with accretion-disk photons, resulting in an estimated muon-neutrino event rate of $\approx 0.3~\mathrm{yr}^{-1}$ during the flaring state with the flux peaking at $\sim1$ PeV. Future observations of TeV $\gamma$-rays by CTA and LHAASO will further constrain cosmic-ray production in this source. 
\end{abstract}




\section{Introduction}

Recent multimessenger studies have reported growing evidence for a statistical association between high-energy neutrinos and active galactic nuclei (AGN), particularly blazars \citep[see, e.g.,][]{Fermi-LAT:2019hte, Giommi:2020hbx, 2020PhRvL.124e1103A, IceCube:2023htm, 2024A&A...681A.119R}. This motivates renewed efforts to identify their contribution to the diffuse TeV–PeV astrophysical neutrino flux discovered by IceCube at $\gtrsim 5\sigma$ statistical significance \citep{IceCube:2013low, IceCube:2014stg}. Blazars are a subclass of radio-loud AGN characterized by a relativistic jet pointing towards Earth and powered by accretion onto a supermassive black hole \citep[see, e.g.,][]{1995PASP..107..803U}. The $\gamma$-ray-emitting AGNs have long been considered potential candidate sources of cosmic rays and hence, neutrinos \citep{Eichler_1979, Sikora_1987, Berezinskii_1989, Stecker_1991, Mannheim_1992, Szabo_1994, Atoyan_2001, Becker_2008, Murase_2014, Petropoulou_2015, Padovani:2016wwn, Palladino_2019, Yuan_2020}, and also ultrahigh-energy cosmic rays \citep[$E\gtrsim10^{17}$ eV;][]{2010APh....33...81E, Murase_2012, Razzaque_2012, Tavecchio:2013fwa, Rodrigues:2020pli, Mondal:2022zvv, Das:2019gtu, Das:2025tfq, Das:2025vqd}. Since the first spatial and temporal association of a $\approx0.3$ PeV $\nu_\mu$ track IC-170922A with a flaring $\gamma$-ray blazar TXS~0506+056 \citep{2018Sci...361..147I, 2018Sci...361.1378I, MAGIC:2018sak}, subsequent studies have explored possible correlations extending to emissions in radio and optical wavebands \citep{Hovatta:2020lor, Plavin:2020mkf, Plavin:2022oyy, Kouch:2024xtd}.
The non-thermal emission from blazars spans radio to very-high-energy (VHE; $\epsilon_\gamma\gtrsim0.1$ TeV) $\gamma$-rays and exhibits variability over minutes to days timescales, which may be uncorrelated in individual wavebands \citep{1996ASPC..110..391U, 2008bves.confE..14M, 2010PhDT.......398B, 2011IAUS..275..164F, 2013MNRAS.431.1618B, 2020NatCo..11.4176S, 2020ApJ...891..120B, 2024ApJS..270...22W, 2024A&A...686A.228O, 2025A&A...703A.259V, 2025ApJ...984...45X}.

Blazars can be further classified based on optical features, e.g., flat-spectrum radio quasars (FSRQs), which exhibit strong and broad optical emission lines, with equivalent width ($|W_{\lambda}|>$\,\text{5\AA}), and BL Lacertae objects (BL Lacs), which have weak ($|W_{\lambda}|<$\,\text{5\AA}) or absent emission lines \citep{1991ApJ...374..431S, 2007Ap&SS.309...63P}. The broadband spectral energy distribution (SED) of blazars consists of two peaks. A low-energy peak between the optical and X-ray energy regimes is believed to arise from synchrotron radiation emitted by relativistic electrons in the magnetic field of the jet. Under a leptonic model, the high-energy $\gamma$-ray peak can originate from synchrotron self-Compton (SSC) emission or inverse-Compton (IC) scattering of external photon fields such as the disk \citep{1993ApJ...416..458D}, broad-line regions \citep[BLR;][]{1994ApJ...421..153S}, or dusty torus \citep{2000ApJ...545..107B}. In the hadronic scenario, external photons are Doppler-boosted in the comoving jet frame and serve as target photons for IC emission and photohadronic interactions ($p\gamma$).
Relativistic protons accelerated in the jet can undergo $p\gamma$ interactions with synchrotron and external photons, or $pp$ interactions with cold protons, resulting in secondary cascade radiation, and neutrinos \citep{1989A&A...221..211M}. This hybrid model has been successful in explaining VHE emission from blazars and in studying blazar-neutrino correlations \citep{Mastichiadis:2005nj, 2012A&A...546A.120D, 2013ApJ...768...54B, Petropoulou:2014rla, Cerruti:2014iwa, 2024A&A...683A.225S, 2024ApJ...971..146X, Prince:2023bhj, Klinger:2023zzv, 2025ApJ...986..110J, Lian:2026gud, Khatoon_2026}, in particular for the IC-170922A event \citep[see, e.g.,][]{Keivani:2018rnh, Cerruti:2018tmc, Liu:2018utd, 2020PASJ...72...20C, 2020PhRvD.101f3024B, Zhang:2019htg, Sahu:2020eep, 2022MNRAS.509.2102G, Das:2022nyp}. It has also been extended to incorporate two emission zones \citep{Ghisellini_2005, Aleksic_2011, Prince:2019xxh, Xue:2019txw}. Alternatively, ultrarelativistic protons can emit proton-synchrotron radiation in a strong magnetic field (approximately 10–100 G) and contribute to the high-energy peak \citep[e.g.,][]{2000NewA....5..377A, Aharonian:2001dm, 2003APh....18..593M, Petropoulou_2012, 2013MNRAS.434.2684M, 2017MNRAS.467L..16P, 2023PhRvD.107j3019X, 2024ApJS..271...10W, Sahakyan:2025phb}.

In this article, we investigate the first-ever multiwavelength flaring detection of the FSRQ PKS 1725+123 starting from early 2025 onward. The $\gamma$-ray emission was found to extend up to $\gtrsim1$ TeV energies and exhibits a hard, unattenuated spectrum. The {\it Fermi} Large Area Telescope ({\it Fermi}-LAT) detected $\gamma$-ray flux enhancements by more than 60 times relative to the average flux reported in  4FGL-DR4 \citep{2025ATel17094....1M, 2025ATel17316....1L, 2025ATel17370....1V}. Similar enhancements were also observed in optical (4-5 orders of change in magnitude) \citep{2025ATel17345....1G}, mm/cm-radio (3-5 times) \citep{2025ATel17376....1S, 2025ATel17356....1H} and the NIR flare (52 times) obtained in the JHK filters (J=13.971, H=12.917, K=11.953) \citep{2025ATel17127....1C, 2025ATel17415....1C}, which is brighter than the previous NIR flares reported in \cite{2012ATel.4201....1C} and \cite{2014ATel.6240....1C}. Furthermore, in this high flux state, high polarization features are noted in both the optical (35\%) and radio wavebands (5\%) \citep{2025ATel17393....1B, 2025ATel17356....1H}.
Additionally, the source was detected for the first time in the VHE regime using H.E.S.S. (6$\sigma$ significance), MAGIC and LST-1 \citep[5$\sigma$ significance;][]{2025ATel17344....1P, 2025ATel17346....1W}. This simultaneous flaring in radio, optical/NIR to GeV–TeV bands positions PKS 1725+123 as an extreme TeV blazar. It is also a candidate counterpart to the IceCube Astrotrack Bronze alert IC201021A, with an angular separation of $2.56^\circ$ and lying within the 90\% containment region \citep[see.,][]{GCN28715, 2022A&A...663A.129N, 2020ATel14111....1B, 2020ATel14192....1G}. This connection suggests that flaring activity may be associated with the efficient acceleration of electrons and cosmic ray protons. In this work, we model the flare origin in light of one-zone leptohadronic emission and estimate the neutrino production rate during this episode.

In Sec.~\ref{DATA}, we describe the details of the multiwavelength data reduction of {\it Fermi}-LAT, {\it Swift}-XRT and {\it Swift}-UVOT instruments. In Sec.~\ref{sec:results} we present the broadband SED and our multiwavelength modelling, including the expected neutrino event rates in IceCube and its connection to activity in different bands. We discuss our results and the implications for cosmic-ray acceleration and multiwavelength emission in Sec.~\ref{sec:discussions}. We present the details of our numerical methods for hadronic cascade in the comoving jet frame in Appendix~\ref{app:hadronic}.

\section{Data Analysis} \label{DATA}

\subsection{{\it Fermi}-LAT Data Analysis}
\begin{figure*}
    \centering
    \includegraphics[width=0.9\textwidth]{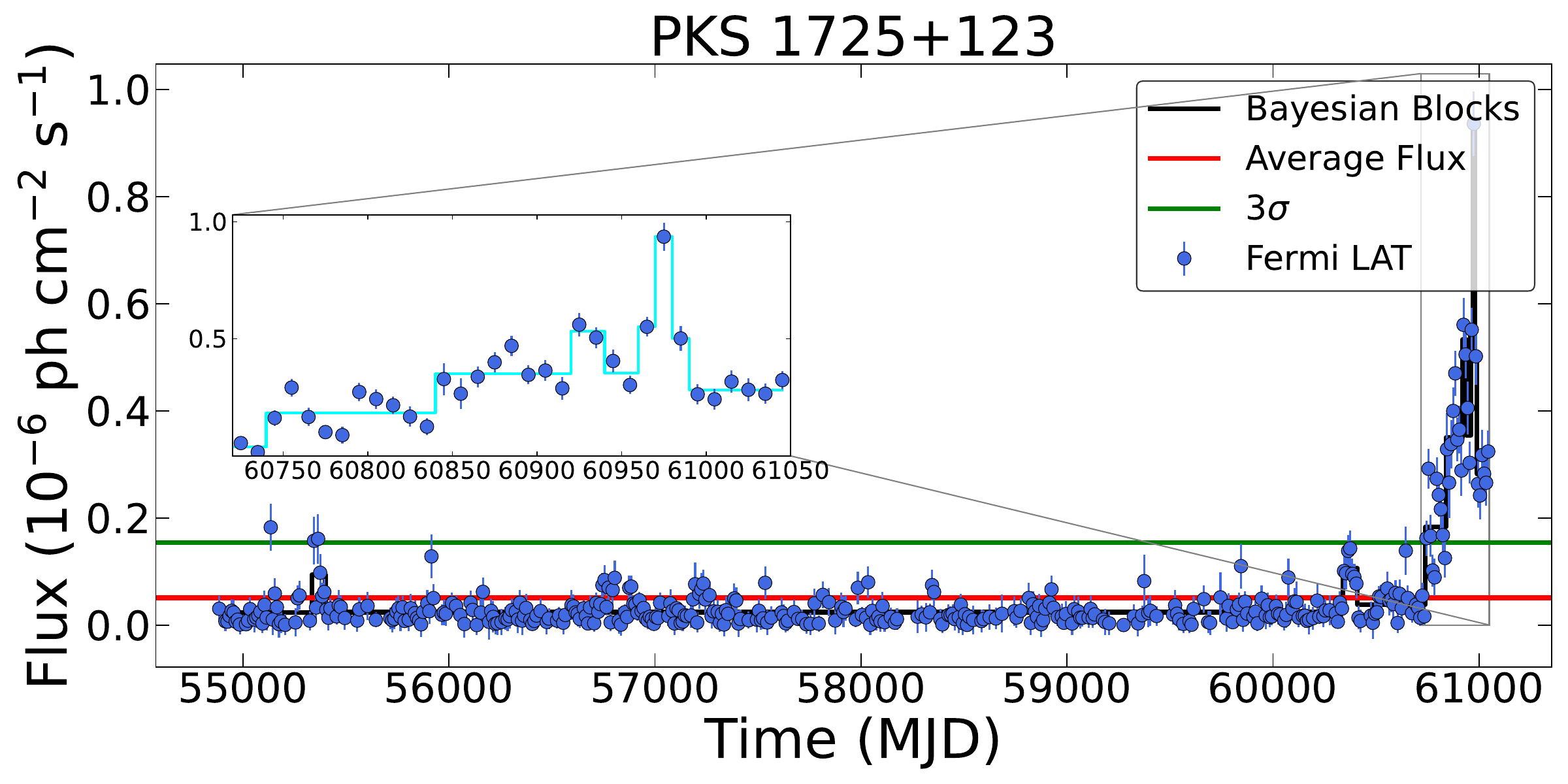}
    \caption{The long-term {\it Fermi}-LAT $\gamma$-ray light curve of PKS 1725+123, binned over 10 days, with Bayesian Blocks (black). Blue points show flux measurements with $TS>1$. Horizontal lines indicate the mean flux (red) and three times the mean (green). The inset shows a zoomed view of the interval MJD 60745-61050.}
    \label{Full Fermi-LAT lightcurve}
\end{figure*}

We utilized $\gamma$-ray data from the {\it Fermi} Large Area Telescope (LAT) in the 0.1-300 GeV energy range for an area of $<15^{\circ}$ around the source. We have created a long-term {\it Fermi}-LAT light curve spanning 2009 to 2025 with 10-day binning to observe the long-term trend as depicted in the Figure \ref{Full Fermi-LAT lightcurve}. Additionally, we have created the 1-day binned lightcurve covering the flaring cascade period from March 2025 to October 2025 (MJD: 60735 to 60985) as shown in the top panel of Figure \ref{fig:bb_lightcurve}, covering flaring episodes. We have processed the data using standard {\tt Fermipy v1.2} software \footnote{\url{https://Fermipy.readthedocs.io/en/latest/}} pipeline, to generate light curve and spectrum \citep{Wood:2017TJ, 2009ApJ...697.1071A}. We selected events of class SOURCE (evclass = 128, evtype = 3) within a $<15^{\circ}$ radius region of interest (ROI) centered at the source. To ensure data quality, we applied the standard zenith angle cut ($<90^{\circ}$) and filter (\texttt{DATA\_{QUAL}>0 \&\& LAT\_{CONFIG}==1}) to establish good time intervals (GTI). The source model included all cataloged objects from the {\it Fermi}-LAT Fourth Source Catalog within the ROI, along with the Galactic diffuse emission model (\texttt{gll\_iem\_v07.fits}) and instrument response function (IRF) (\texttt{iso\_P8R3\_SOURCE\_V3\_v1.txt}) for background modelling. We performed the following steps to acquire the average $\gamma$-ray flux of the source. After an initial optimization with the {\tt Fermipy} optimize tool, the normalizations and spectral parameters of all sources within $<15^{\circ}$ of the ROI center were allowed to vary freely. The significance of source emission was evaluated using the test statistic ($TS = 2\log_{10}\left({L_0/L_1}\right)$
, \cite{1996ApJ...461..396M}), which is defined as corresponding to the probabilities of the null hypothesis (no source) and the alternative hypothesis (with source), respectively. All spectral parameters were kept fixed for sources with TS$<$9, and sources with negative TS values were excluded from the analysis.

\subsection{Swift Data Analysis}
The {\it Swift} telescope has observed this source multiple times from March 2025 to November 2025, covering flaring activities. The X-ray spectra and light curve were extracted in the energy range of 0.3-10 keV using {\texttt{Swift-XRT data products generator}} \citep{2009MNRAS.397.1177E}. The X-ray lightcurve in the 0.3-10 keV covering the flare period (MJD: 60735 to 60985) observed by {Swift}-XRT instrument is displayed in the second top panel of Figure \ref{fig:bb_lightcurve}. The spectral fitting was performed by {\texttt{XSPEC}} \footnote{\url{https://heasarc.gsfc.nasa.gov/docs/software/xspec/}}, keeping all parameters free. The X-ray spectra are shown in \ref{XRT_spectra}, with column density $N_H = 9\times10^{20}\rm cm^{-2}$, close to its nominal values.
{\it Swift}-UVOT, which has three optical (V, B, U) and three ultraviolet (W1, M2, W2) filters, offers complementary multiwavelength coverage to the XRT. We have processed the data using the standard UVOT pipeline\footnote{\url{https://www.swift.ac.uk/analysis/uvot/}}, with the {\texttt{HEASoft v6.32}} and {\texttt{CALDB}}. A 5\arcsec and 10\arcsec circular regions were used for the source and background count estimation, respectively. The {\texttt{UVOTIMSUM}} command is used to sum data within the specified intervals for each filter, and task {\texttt{UVOTSOURCE}} was used to estimate the source magnitudes in the respective filters. The resultant magnitudes were also corrected for Galactic extinction as suggested in \cite{2011ApJ...737..103S} before converting into fluxes using the flux conversion factors of \citet{2016MNRAS.461.3047L} and the zero points from \cite{2011AIPC.1358..373B}.

\section{Results}\label{sec:results}
\subsection{Multi-wavelength Light Curves}
\begin{figure}
    \centering
    \includegraphics[angle=0,width=0.98\linewidth]{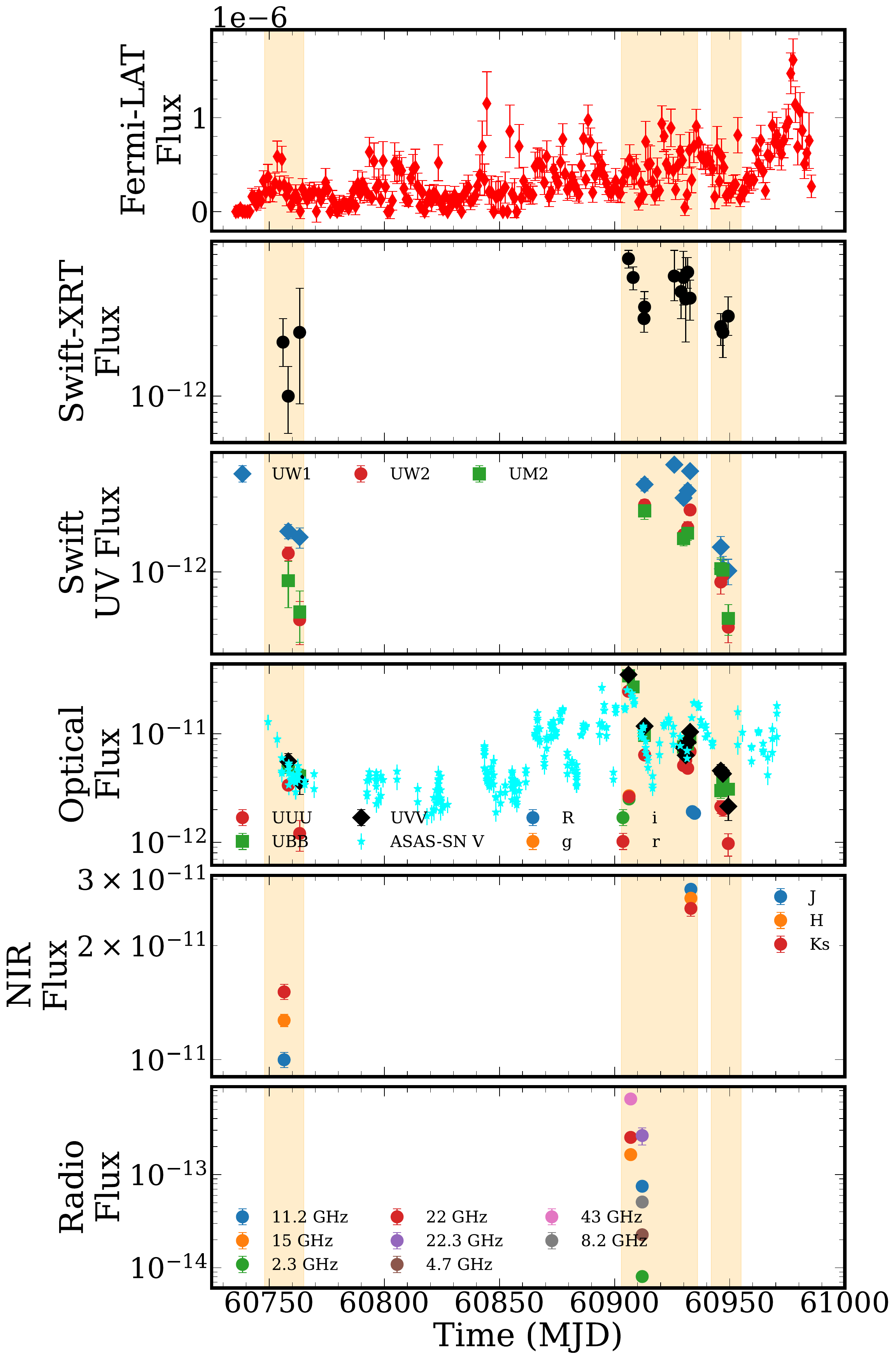}
    \caption{Multiwavelength light curve of PKS 1725+123 during its high activity state starting from March 2025. The shaded regions (MJD1: 60748-60765, MJD: 60903-60936, MJD3: 60942-60955) show the three flaring events selected in this study. Here the one-day binned {\it Fermi}-LAT fluxes are reported in $\mathrm{ph\,cm^{-2}\,s^{-1}}$
, and X-ray to Radio fluxes expressed in $\mathrm{erg\,cm^{-2}\,s^{-1}}$.}
    \label{fig:bb_lightcurve}
\end{figure}

The long-term {\it Fermi}-LAT lightcurve of PKS 1725+123 is presented in Figure \ref{Full Fermi-LAT lightcurve}, showing that the source was in a low-activity state from 2009 to 2024. In 2025, the source transitions from a low-flux to a high-flux state, with the flux rising to 10$^{-6}$ ph cm$^{-2}$ s$^{-1}$, well above the historical average. The flaring part is zoomed in on the inset plot and fitted with a Bayesian block to represent the flux rising. In Fig. \ref{fig:bb_lightcurve}, we show the first-ever broadband variation of the source observed during the high-activity phase in 2025. We have estimated its flux variability in optical/UV, X-ray, and $\gamma$-ray bands using several methods discussed below such as Variability Amplitude (\text{VA}), Fractional Variability (\text{$F_{\rm var}$}), and minimum variability timescale (\text{$\tau_{\rm var}$}) for the MWL lightcurves presented in Figure \ref{fig:bb_lightcurve}. The estimated variability properties are given in the Table \ref{UVOT_RESULTS}.

The \text{VA} is used to characterize the relative flare amplitudes, is defined as follows: 
\begin{equation}
    VA = \frac{F_{\mathrm{max}}-F_{\mathrm{min}}}{F_{\mathrm{min}}}, \label{eqn:VA}
\end{equation}
where \text{$F_{\rm min}$} \text{$F_{\rm max}$} are the minimum and maximum flux values in a given LC. The associated errors are \text{VA} is estimated using error propagation formulae \citep[][]{2003drea.book.....B}:
\begin{equation}
    \sigma_{\mathrm{VA}} = (VA+1)\cdot\sqrt{\left(\frac{\sigma_{F_{\mathrm{max}}}}{F_{\mathrm{max}}}\right)^2+\left(\frac{\sigma_{F_{\mathrm{min}}}}{F_{\mathrm{min}}}\right)^2},
\end{equation}

The estimated \text{VA} indicates that the $\gamma$-ray LC shows the highest amplitude with \text{VA}$\sim$50, signifying intense flaring activity that corresponds with the values mentioned in \citep{2025ATel17316....1L}. Among the {\it Swift}-UVOT filters, the U band exhibits the largest variations with \text{VA}$\sim$25, followed by moderate variability in the V and B bands (around $\sim$15 and $\sim$10), while significantly lower amplitudes in the UV filters W1, M2, and W2 (approximately $\sim$3-5) suggest that variability decreases as the UV wavelengths shorten. This demonstrates how significantly the \text{VA} values are influenced by wavelength and energy. A minimal variability in the X-ray range is indicated by {\it Swift}-XRT (\text{VA}$\sim$6), which is similar to the lower-UV measurements.


\begin{table*}
    \centering
    \caption{UVOT variability results: Col. 1: Instrument; Col. 2: UVOT filters; Col. 3: Mean flux (\ergcm{-12}); Col. 4 Fractional Variability ($F_{var}$); Col. 5: Variability Amplitude (VA); Col. 6: Minimum Variability Timescale ($\tau_{\rm var}$).}
    \begin{tabular}{cccccccc}
        \hline
        Instrument & Filter & $<F>$ & $F_{\mathrm{var}} (\%)$ & VA & $\tau_{\mathrm{var}} (days)$ &  \\ 
        \hline
      Swift-UVOT  & V & 9.078$\pm$0.390 & 100.483$\pm$1.332 & 15.387$\pm$2.170 & 12.232$\pm$0.005 &  \\
         & B & 10.224$\pm$0.287 & 97.491$\pm$0.838 & 10.448$\pm$0.892 & 18.157$\pm$0.005 &  \\
         & U & 5.751$\pm$0.182 & 115.066$\pm$1.001 & 24.349$\pm$2.992 & 19.293$\pm$0.003 &  \\
         & W1 & 2.560$\pm$0.122 & 53.325$\pm$1.521 & 3.739$\pm$0.449 & 11.367$\pm$0.008 &  \\
         & W2 & 1.427$\pm$0.075 & 57.113$\pm$1.791 & 5.006$\pm$0.664 & 12.269$\pm$0.005 &  \\
         & M2 & 1.229$\pm$0.098 & 53.429$\pm$2.956 & 3.837$\pm$0.605 & 17.829$\pm$0.009 &  \\  \\
      Swift-XRT & - & 3.696$\pm$0.363 &  29.107$\pm$8.580 &  5.999$\pm$2.758  & 118.569$\pm$12.229 & \\  
      {\it Fermi}-LAT & - & 3.449$\pm$0.171 &  68.151$\pm$2.686 & 50.597$\pm$0.324 & 0.250$\pm$0.001 & \\  
        \hline
    \end{tabular}
    \label{UVOT_RESULTS}
\end{table*}

The \text{VA} allows us to detect the large amplitude variations, thus, to quantify the intrinsic variability amplitude relative to the mean, we calculated the \text{$F_{\rm var}$} \citep[see.,][]{1990ApJ...359...86E, 2003MNRAS.345.1271V}. For a given LC with N data points, with a sample variance of ($S^2$), a mean square uncertainty in flux ($\langle\sigma^2_{\rm err}\rangle$), and with a mean flux ($\langle F \rangle$), the $F_{\rm var}$ is given by:

\begin{equation} \label{FVs}
    F_{\rm var} = \sqrt{\frac{S^{2} - \langle{\sigma^2_{\rm err}}\rangle}{{\langle F \rangle}^2}},
\end{equation}
where the uncertainty in $F_{\rm var}$ is estimated from \citep{2003MNRAS.345.1271V},
\begin{equation}
   \sigma_{F_{\mathrm{var}}} =  \sqrt{\left(\frac{1}{\sqrt{2N}}\frac{\langle{\sigma^2_{\rm err}\rangle}}{F_{\rm var}}\frac{1}{{\langle F \rangle}^2}\right)^2 + \left(\sqrt{\frac{\langle{\sigma^{2}_{\rm err}}\rangle}{N}}\frac{1}{{\langle F \rangle}^2}\right)^2}. 
\end{equation}

Our analysis reveals high variability in the source in 2025, with \text{$F_{\rm var}$} frequency-dependent. For {\it Swift}-UVOT, in the optical band (\text{$F_{\rm var}\sim$}115\%, 100\%, and 97\% for U, V, and B-bands, respectively), the source shows two times more variability than UV (\text{$F_{\rm var}\sim$}53\%, 57\%, and 53\% for W1, W2, and M2-bands, respectively). On the other hand, its X-ray variability decreases drastically to about 30\%. However, the 1-day binned {\it Fermi}-LAT LC once again shows high \text{$F_{\rm var}\sim$}68\%.

The minimum-variability timescale $t_{\rm var}$ is a crucial tool for constraining the sizes of the emitting regions. For blazars, rapid flux changes over timescales of minutes to hours to days suggest that the regions causing these variations lie near the supermassive black hole. The expression for $t_{\rm var}$ is defined as \citep[see.,][]{1974ApJ...193...43B},
\begin{equation} \label{tvar}
    \tau_{\mathrm{var}} = \left|\frac{\Delta t}{\Delta \ln F}\right|,
\end{equation}
where $\Delta t = |t_2 - t_1|$ and $F_1$ and $F_2$ stand for the flux values at times $t_1$ and $t_2$, respectively. Only flux pairs that satisfied the following requirements were considered: (i) $F_2 > F_1$, and (ii) $F_2-F_1 > 3\sqrt{\sigma^2_{F_1} + \sigma^2_{F_2}}$, where $\sigma_{F_1}$ and $\sigma_{F_2}$ indicate the corresponding measurement uncertainties. The uncertainty in the \text{$\tau_{\rm var}$} is derived using error propagation formulae as given in \cite{2008ApJ...672...40H}. The uncertainty in the variability timescale is given by a generalized error propagation formula akin to Eq. (3.14) in \citet{2003drea.book.....B}:
\begin{equation} 
    \Delta_{\tau_{\mathrm{var}}} \approx \sqrt{\frac{F_1^2\Delta F_2^2+F_2^2\Delta F_1^2}{F_1^2 F_2^2 (\ln[F_1/F_2])^4}} \Delta t,
\end{equation}
For the UVOT filters, the \text{$\tau_{\rm var}$} varies from 11\text{ks} to 19\text{ks}, while the X-ray analysis suggests a higher \text{$\tau_{\rm var}\sim$118 ks}, while the $\gamma$-ray analysis suggests the fastest variability of $\leq$6 hours. The \text{$\tau_{\rm var}$} can be used to constrain the radius of the emitting region,
\begin{equation}
     R^\prime \lesssim \frac{\delta_D c\tau_{\rm var}}{1+z} = 1.94\times10^{16}\, \text{cm} \times(1+z)^{-1}\left(\frac{\delta_D}{30}\right)\left(\frac{\tau_{\rm var}}{6\, \text{hrs}}\right) 
      \label{eqn:radius}
\end{equation}
where $\delta_D$ is the Doppler factor. In our spectral modeling in Sec.~\ref{subsec:sed_modeling}, we show $\delta_D \simeq 30$ well explains the data during flaring episodes, consistent with the range for Fermi-detected FSRQs \citep{Pushkarev_2009, Liodakis:2018kmv}. 
Additionally, \text{$\tau_{\rm var}$} value can also be used to determine the location of the emitting region ($R_H$) relative to the central black hole, assuming
$R_H \simeq {R^\prime}/{\theta_{\rm ob}}$
where $\theta_{\rm ob}$ is the viewing angle with respect to the jet axis \citep[see, e.g.,][]{Tavecchio_2010}. Using the 
approximations $\delta_D \simeq \Gamma$ and $\Gamma \theta_{\rm ob} \simeq 1$, which holds for blazars, since the jets are highly collimated and aligned towards us \citep[$\theta_{\rm ob}\leq5^{\circ},$][]{1995PASP..107..803U}, we obtain
\begin{equation}
    R_H \lesssim \frac{2 c \tau_{\rm var} \delta^2_D}{1 + z}= 1.17\times10^{18}\, \text{cm} \times (1+z)^{-1} \left(\frac{\delta_D}{30}\right)^2\left(\frac{\rm \tau_{\rm var}}{6\, \text{hr}}\right)
    \label{eqn:location}
\end{equation}
It should also be noted that Eqns.~\ref{eqn:radius} and \ref{eqn:location} are approximate and based on the upper limit to the variability timescale.

\subsection{Broadband SED modelling \label{subsec:sed_modeling}}

\subsubsection{Model Setup}

We model the multiwavelength SEDs for the three flaring epochs using a one-zone leptohadronic model. The emitting region (distinct from the acceleration site) is a spherical blob of comoving radius $R'$ containing a relativistic electron--proton plasma permeated by a uniform magnetic field $B'$. Here, the primed quantities correspond to those in the comoving jet frame. 
The bulk Lorentz factor of the plasma is $\Gamma$ and Doppler factor $\delta_D=[\Gamma(1-\beta\cos\theta_{\rm ob})]^{-1}$, where $\beta c$ is the plasma speed. For highly beamed emission, we consider $\theta_{\rm ob}\lesssim 1/\Gamma$, and adopt $\delta_D\simeq\Gamma$.
We also assume that the electrons follow a power-law spectrum and are injected into the blob,
\begin{align}
    Q'_e(\gamma'_e) = A_e \left( {\gamma'_e}/{\gamma_0} \right)^{-\alpha} \quad \text{for } \gamma'_{e, \min} < \gamma'_e < \gamma'_{e, \max} \, .
\end{align}
$A_e$ is fixed by the injected electron luminosity and $\gamma_0 m_ec^2=0.5~\mathrm{GeV}$ is a reference energy. A quasi-steady state is reached when injection is balanced by radiative cooling and/or escape, such that $N'_e(\gamma'_e) = Q'_e(\gamma'_e)\, t'_e$
with $t'_e = \min(t'_{\rm cool}, t'_{\rm esc})$, and $t'_{\rm esc} \simeq t'_{\rm dyn} = 2R'/c$. The radiative cooling time, for synchrotron and IC emission, is
\begin{align}
    t'_{\rm cool} = {3 m_e c}/{4 \sigma_T \gamma'_e (u'_B + \kappa_{\rm KN} u'_{\rm ph})},
\end{align}
where $u'_B=B'^2/8\pi$ is the energy density in the magnetic field, $u'_{\rm ph}$ is the soft-photon energy density, and $\kappa_{\rm KN}$ accounts for Klein--Nishina suppression of IC scattering.
We solve the  transport equation using the open-source code \textsc{GAMERA} \citep{Hahn:2015hhw, Hahn_2022} to find the steady-state solution to
\begin{align}
    \frac{\partial N'_e}{\partial t} = Q'_e(\gamma'_e,t') - \frac{\partial}{\partial \gamma'_e}(b N'_e) - \frac{N'_e}{t_{\rm esc}},
\end{align}
where $b(\gamma'_e, t') = |\dot\gamma^\prime_e|$ is the energy loss rate.
The steady-state electron spectrum produces synchrotron (SYN), synchrotron self-Compton (SSC) emission, and external inverse-Compton (IC) scattering of blackbody photon fields arising from the broad line region and the accretion disk, both required to fit the observed SED. 




We follow the numerical methods presented in \citet{Das:2022nyp, Prince:2023bhj} to model the hadronic cascade from $p\gamma$ interactions and extend it to include inverse-Compton cooling of secondary electrons (see Appendix~\ref{app:hadronic}). Protons are injected with a power-law spectrum $dN/dE_p=A_p E_p^{-\alpha}$ and the minimum proton energy is set to $\gamma^\prime_{p, \rm min} = 10$, typical for hadronic emission models \citep{Gao:2018mnu}. The cutoff energy in the jet is determined from SED modelling. Their dominant energy losses arise from photopion production ($p\gamma\rightarrow p+\pi^0$ or $n+\pi^+$) and Bethe--Heitler (BH) pair production ($p\gamma\rightarrow p+e^+e^-$), with target photons provided by the leptonic radiation and the external photon fields. 
Charged pion decays yield neutrinos. The normalization $A_p$ is fixed by the required proton kinetic power. Secondary $\gamma$ rays and $e^\pm$ from pion decay, as well as $e^\pm$ pairs from the BH channel, produce cascade radiation. We take into account $\gamma\gamma\rightarrow e^\pm$ absorption to calculate the steady state spectrum of secondary electrons (see Appendix~\ref{app:hadronic}). For a distant source like PKS~1725+123 at a redshift of $z=0.586$, high-energy $\gamma$-rays are absorbed in the extragalactic background light (EBL) consisting of infrared, optical, and UV photons. We use the \citet{Gilmore_2012} EBL model to incorporate the attenuation at high energies.

For proton energies below a few PeV, the interaction time $t^\prime_{p\gamma}$ typically exceeds the dynamical time $t'_{\rm dyn}\sim R/c$. 
The proton escape time depends on the diffusion coefficient as $t^\prime_{p,\rm esc}\sim R^{\prime 2}/D(E)$, and we parameterize it to satisfy $t^\prime_{p,\rm esc}\gtrsim t^\prime_{p\gamma}$ under a Bohm-like diffusion, $D(E)\propto E$ (see Appendix~\ref{app:hadronic}). A cutoff in the proton spectrum in the jet beyond a maximum energy $\gamma'_{p,\max}\sim 10^6$ is required to explain the SED, which also limits the peak energy of the neutrino spectrum, as in the case of TXS 0506+056 \citep[see, e.g.,][]{Keivani:2018rnh, Gao:2018mnu}. We assume that escape dominates at higher energies, thereby suppressing $p\gamma$ interactions within the jet. Using the Hillas condition \citep{1984ARA&A..22..425H}, the maximum acceleration energy is $E^{\prime \rm \,acc}_{p, \rm max}= 2\beta cZeB^\prime R^\prime \approx 1.2\times10^{19}$ eV for $B^\prime=2$ G and $R^\prime=10^{16}$ cm (see Table~\ref{tab:sed_param}). 
Depending on the energy, protons are affected differently by the magnetic field. 
If the diffusion coefficient is $D\propto E^q$ with $q\gtrsim 1$, interaction efficiency rapidly declines above tens of PeV.  
In the one-zone model, the efficient escape of ultrahigh-energy cosmic rays (UHECRs; $E\gtrsim10^{17}$ eV) may result from a rigidity-dependent diffusion rate, such as $D(E)\propto E^2$ \citep{Globus:2007bi, Harari:2013pea, Muzio:2021zud}. The muon neutrino ($\nu_\mu +\overline{\nu}_\mu$) flux from $p\gamma$ interactions is given by
\begin{equation}
E_\nu^2 J_\nu = ({1}/{3})  (E_\nu'^2 Q'_{\nu, p\gamma})({V' \delta_D^2 \Gamma^2}/{4\pi d_L^2}),
\end{equation}
where the factor $1/3$ accounts for neutrino oscillations and $Q'_{\nu, p\gamma}$ is the neutrino production rate from charged pion decay.

\subsubsection{Multiwavelength SED fit}

Photons from the accretion disk and the BLR are Doppler-boosted into the jet frame and serve as target photons for IC scattering by relativistic electrons. Their temperatures and energy densities are constrained through the multiwavelength SED fits. The distance of the emission region along the jet determines the dominant target photons for IC scattering. It has been shown that for radio-loud AGNs, particularly quasars, at redshift $z<0.5$, the black hole can have a mass of $\langle M_{\rm BH}/M_\odot\rangle\approx10^9$ \citep{Falomo:2003up}. We extrapolate this result to estimate the Eddington luminosity of the source as, $L_{\rm Edd}\simeq1.3\times10^{38}\,\text{erg s}^{-1}(M_{\rm BH}/M_{\odot})\approx 1.3\times 10^{47}$ erg s$^{-1}$. The magnetic field $B^\prime$, emission-region radius $R^\prime$, and maximum electron Lorentz factor are constrained by fitting the radio and optical data with a synchrotron model. To calculate the IC spectrum, we approximate both external photon fields as quasi-isotropic in the comoving frame, using the temperatures and energy densities inferred from the fit.
For consistency in the acceleration process of both leptons and hadrons, we assume a power-law spectrum with a sharp cutoff and the same spectral index of $\alpha=2$.

The angle-integrated comoving energy density of the direct accretion-disk radiation is given as \citep{Dermer_2009},
\begin{equation}
u^\prime_{\rm disk}(R_H)=\frac{3GM_{\rm BH}\dot m}{8\pi c\,R_H^3\Gamma}\left(2\Gamma^3 c_d+\frac{R_H}{4\Gamma R_{\min}}\right),
\end{equation}
where the first and second terms are the near-field (NF) and far-field (FF) contributions, respectively.
For a Schwarzschild black hole, $R_{\min}=6R_g$, with $R_g=GM_{\rm BH}/c^2\simeq1.48\times10^{14}\,\mathrm{cm}$ for $M_{\rm BH}=10^9\,M_\odot$. We assume the FF contribution is negligible compared to the NF component. Using $L_{\rm disk}=GM_{\rm BH}\dot{m}/2R_{\rm min}$ and $c_d\simeq0.023$ for a standard optically thick Shakura-Sunyaev disk with zero-torque inner boundary condition, and adopting $L_{\rm disk}=10^{46}\,\mathrm{erg\,s^{-1}}$ and $\Gamma\simeq30$, the near-field approximation yields $R_H=1.25\times10^{17}\ \mathrm{cm}$ for the best-fit $u^\prime_{\rm disk}=1.5$ erg cm$^{-3}$ from the SED fit (Table~\ref{tab:sed_param}). The value is comparable to that obtained in Eqn.~\ref{eqn:location} using variability timescale and geometrical approximations.

%

For an isotropic BLR photon field in the AGN frame, the comoving energy density in the jet frame is \citep{Barnacka:2013oxa},
\begin{align}
    u^\prime_{\rm BLR}=\frac{4}{3}\Gamma^2 u_{\rm BLR}=\frac{\xi_{\rm BLR}L_{\rm disk}\Gamma^2}{3\pi R_{\rm BLR}^2c}
\end{align} 
Assuming a BLR reprocessing fraction of $\xi_{\rm BLR}=0.02$ \citep{Prince:2018rxa}, the above expression gives a BLR radius of $R_{\rm BLR}\simeq 1.6\times10^{18}$ cm ($\simeq 0.52$ pc) for the best-fit $u^\prime_{\rm BLR}=0.25$ erg cm$^{-3}$ obtained from the SED fit. 
The contribution from the dusty torus is ignored in our analysis, as it becomes significant only for emission regions at much larger distances, $\gtrsim 1$ pc \citep[see, e.g.,][]{Sikora_2009}.

\begin{table}
\centering
\caption{Parameter values for multiwavelength SED modelling fit results. The non-thermal electron and proton injection spectrum is given by a power-law dN/dE $\propto E^{-2}$. The luminosities are presented in the AGN frame, while primed quantities correspond to the comoving jet frame.}
\label{tab:sed_param}
\renewcommand{\arraystretch}{1.1}
\begin{tabular}{l|ccc}
\hline\hline
 & \multicolumn{3}{c}{Leptohadronic} \\
\cline{2-4}
Parameters & Epoch - 1 & Epoch - 2 & Epoch - 3 \\
\hline
$\alpha$ ($e/p$ spectral index) & \multicolumn{3}{c}{2.0 (fixed)}\\
$E_0$ [MeV] & \multicolumn{3}{c}{500 (fixed)}\\
\hline
$\delta_D$ & \multicolumn{3}{c}{30}\\
$T'_{\rm BLR}$ [K] & \multicolumn{3}{c}{$5\times10^4$ }\\
$u'_{\rm BLR}$ [erg cm$^{-3}$] & \multicolumn{3}{c}{0.25 }\\
$T^\prime_{\rm disk}$ [K] & \multicolumn{3}{c}{$6\times10^6$ } \\
$u^\prime_{\rm disk}$ [erg cm$^{-3}$] & \multicolumn{3}{c}{1.5 } \\
$R^\prime$ & \multicolumn{3}{c}{$1.0\times10^{16}$ }\\
$\gamma'_{p, \rm min}$ & \multicolumn{3}{c}{$10^1$ (fixed) }\\
$\gamma'_{p, \rm max}$ & \multicolumn{3}{c}{$10^6$ }\\
\hline

$B'$ [Gauss] & 2.2 & 2.0 & 2.0\\
$\gamma'_{e,\rm min}$ & $2.0\times10^2$ & $2.5\times10^2$ & $2.0\times10^2$ \\
$\gamma'_{e,\rm max}$ & $2.0\times10^3$ & $2.5\times10^3$ & $2.0\times10^3$\\

\hline
$L_{j,e}$ [erg s$^{-1}$] & $5.5\times10^{44}$ & $1.2\times10^{45}$ & $9.3\times10^{44}$\\
$L_{j,B}$ [erg s$^{-1}$] & $1.6\times10^{45}$ & $1.4\times10^{45}$ & $1.4\times10^{45}$ \\
$L_{j,p}$ [erg s$^{-1}$] & $3.6\times10^{47}$ & $3.6\times10^{47}$ & $3.6\times10^{47}$ \\
\hline
\end{tabular}
\end{table}

\begin{figure}
    \centering
    \subfigure[MJD: 60748-60765]{
    \includegraphics[width=0.44\textwidth]{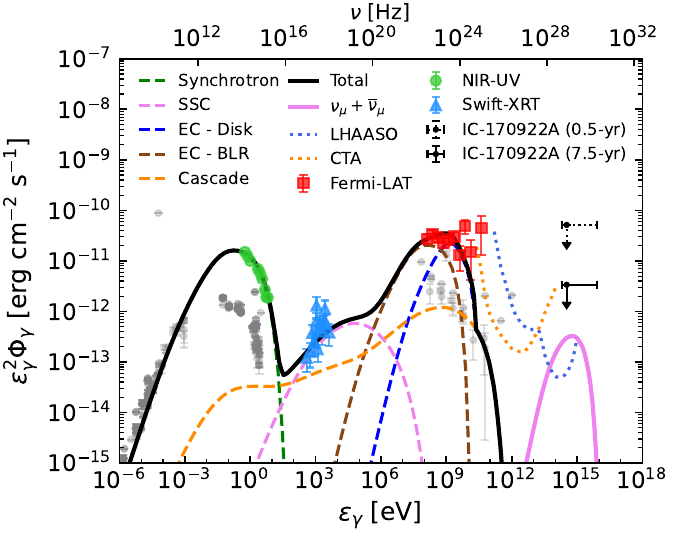}}
    \subfigure[MJD: 60903-60936]{
    \includegraphics[width=0.44\textwidth]{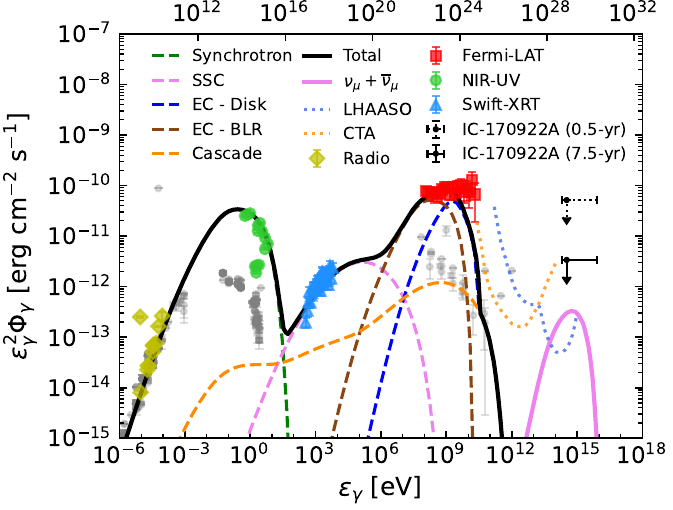}}
    \subfigure[MJD: 60942-60955]{
    \includegraphics[width=0.44\textwidth]{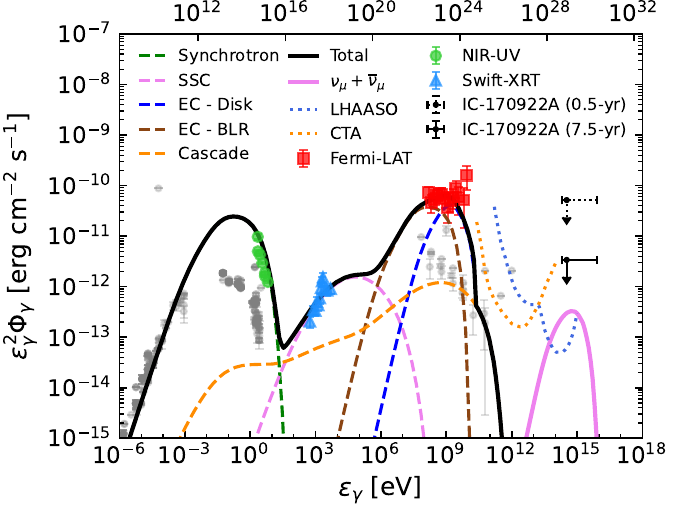}}
    \caption{SED of PKS 1725+123 during the three flaring epochs. The grey data points indicate archival data. The yellow, green, blue, and red data points correspond to the radio, NIR-UV, X-ray, and $\gamma$-ray wavebands, respectively. The black solid line shows the total emission. 
    The black dotted and solid data correspond to the estimated 1-flavor $\nu_\mu+\overline{\nu}_\mu$ flux upper limits from TXS 0506+056 for 1 detection in 0.5 and 7.5 years, respectively \citep{2018Sci...361.1378I}. The orange-dotted line shows the CTAO-N differential point-source sensitivity assuming 50h observation time and pointing to 20 degrees zenith \citep{Gueta:2021vO}. The blue-dotted line is the LHAASO sensitivity to Crab-like $\gamma$-ray point sources for 1-yr observation \citep{Vernetto:2016gro}.}
    \label{fig:mar_flare}
\end{figure}




The best-fit leptohadronic models in the steady state are shown in Fig.~\ref{fig:mar_flare}, together with the corresponding neutrino fluxes. 
Our fits imply a very compact emission region of radius $R^\prime\sim 10^{16}$~cm, placing it close to the base of the jet; consequently, the accretion-disk photon field dominates the IC emission. The required magnetic field remains approximately constant across all epochs, whereas the electron luminosity and energy range vary. Overall, the SEDs in this case are explained predominantly by leptonic emission.

For hadronic interactions, we assume a steady-state cosmic-ray spectrum, and the external photons in the comoving jet frame constitute the main target photons for $p\gamma$. Since lower-energy photons require higher-energy protons to satisfy the $\Delta$-resonance condition, we choose the cutoff proton Lorentz factor, $\gamma_p^\prime$, such that the resulting hadronic emission does not overproduce the X-ray flux. In the comoving jet frame, the resonance condition is $E_p^\prime \epsilon_\gamma^\prime \simeq 0.32(1-\cos\theta_i)^{-1}\,\mathrm{GeV}^2$ for collision at an angle $\theta_{i}$. For an isotropic target photon field, $\langle 1-\cos\theta_i \rangle \sim 1$. The differential photon density in the comoving frame peaks at $\approx 17.2$~eV for the BLR field and $\approx 2.05$~keV for the accretion-disk field. These correspond to proton Lorentz factors of $\gamma_p^\prime \approx 2\times10^7$ and $\approx 1.7\times10^5$, respectively, for $\Delta$-resonance. To suppress photopion interactions with BLR photons, we adopt $\gamma_{p,\rm max}^\prime = 10^6$. Hence, in our modeling, the dominant photon field for hadronic interactions is the accretion disk photons. The normalization of the hadronic component is obtained by maximizing the neutrino flux without overestimating the X-ray flux.

The jet luminosities in electrons ($L_{j,e}$), magnetic field ($L_{j,B}$), and protons ($L_{j,p}$) in the AGN frame are listed in Table~\ref{tab:sed_param}. The inferred proton luminosity, $L_{j,p}$, stays constant over the flaring epochs and is comparable to the Eddington luminosity of the source, which further constrains the neutrino flux in our model. However, we note that the highest-energy \textit{Fermi}-LAT data points are not covered in our one-zone modeling, which shows a rising trend. A multi-zone emission model may be required, with the $\gamma$-ray and neutrino fluxes originating from a distinct region. In this work, we use the one-zone model as the most constraining scenario to calculate the neutrino flux rate. Moreover, H.E.S.S and MAGIC have detected VHE $\gamma$-ray emission extending to multi-TeV energies. A steady flux at this energy range would imply cosmogenic $\gamma$-rays from UHECR interactions \citep[see, e.g.,][]{Das:2022nyp, Prince:2023bhj}, since leptonic emission from the source is significantly absorbed by the EBL.

\subsection{Neutrino event rate}

The expected neutrino flux obtained in all three flaring epochs shows almost no variation due to a steady cosmic-ray spectrum with no luminosity variation over the flaring episodes. Moreover, since the dominant target photons are from the external photon field rather than leptonic emission, the resulting neutrino flux across the individual epochs shows negligible variation. The single flavor $\nu_\mu+\overline{\nu}_\mu$ flux peaks at an energy $\approx 1$ PeV and is shown in Fig.~\ref{fig:mar_flare}.
The $\nu_\mu+\overline{\nu}_\mu$ event rate at IceCube in a given operation time  $\Delta T$ is calculated using the expression
\begin{equation}
    \mathcal{N}_{\nu_\mu} =\Delta T \int_{\epsilon_{\nu, \rm min}}^{\epsilon_{\nu, \rm max}} d\epsilon_\nu \dfrac{d\Phi_\nu}{d\epsilon_\nu} \langle A_{\rm eff} \rangle_\theta
\end{equation}
where $\langle A_{\rm eff} \rangle_\theta$ is the effective detector area averaged over the zenith angle bin concerned \citep{IceCube:2018ndw}. For PKS~1725+123 at a position RA, Dec = (262.03, 12.26), we use the effective area for event selection as a function of neutrino energy and zenith angle at IceCube from \cite{Stettner:2019tok}. We calculate the expected number of muon neutrino events from this source to be $\approx 0.3\times (\Delta T/1\,\text{yr})$, which is much less than that estimated for the blazar TXS~0506+056. Our analysis is restricted to the cosmic-ray luminosity during the flaring period. The expected number of neutrinos covering the entire flaring period of MJD: 60748 - 60955 can be estimated to be $\approx0.2$ track-like events. However, if the cosmic-ray luminosity remains steady beyond this period, a detection is expected over 10 years of IceCube observations.

\begin{figure}
    \centering
    \subfigure[$\gamma$-ray: Flux vs. Spectral Index]{\includegraphics[width=0.99\linewidth]{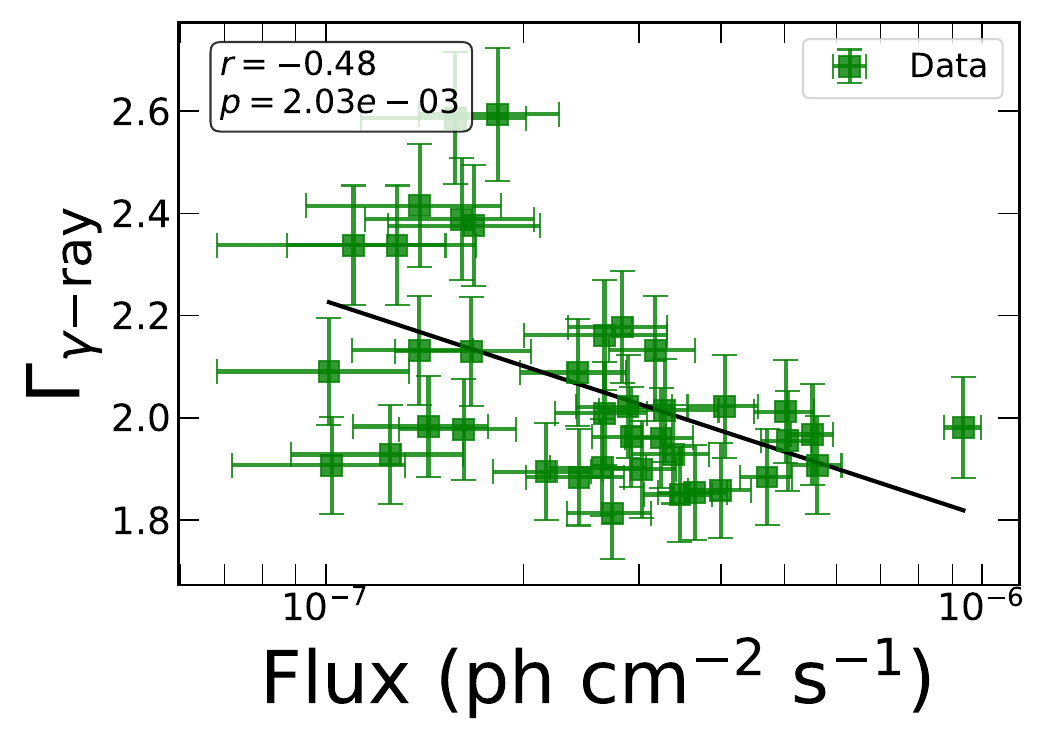} \label{fig:4a}}
    \subfigure[X-ray: Flux vs. Spectral Index]{\includegraphics[width=0.99\linewidth]{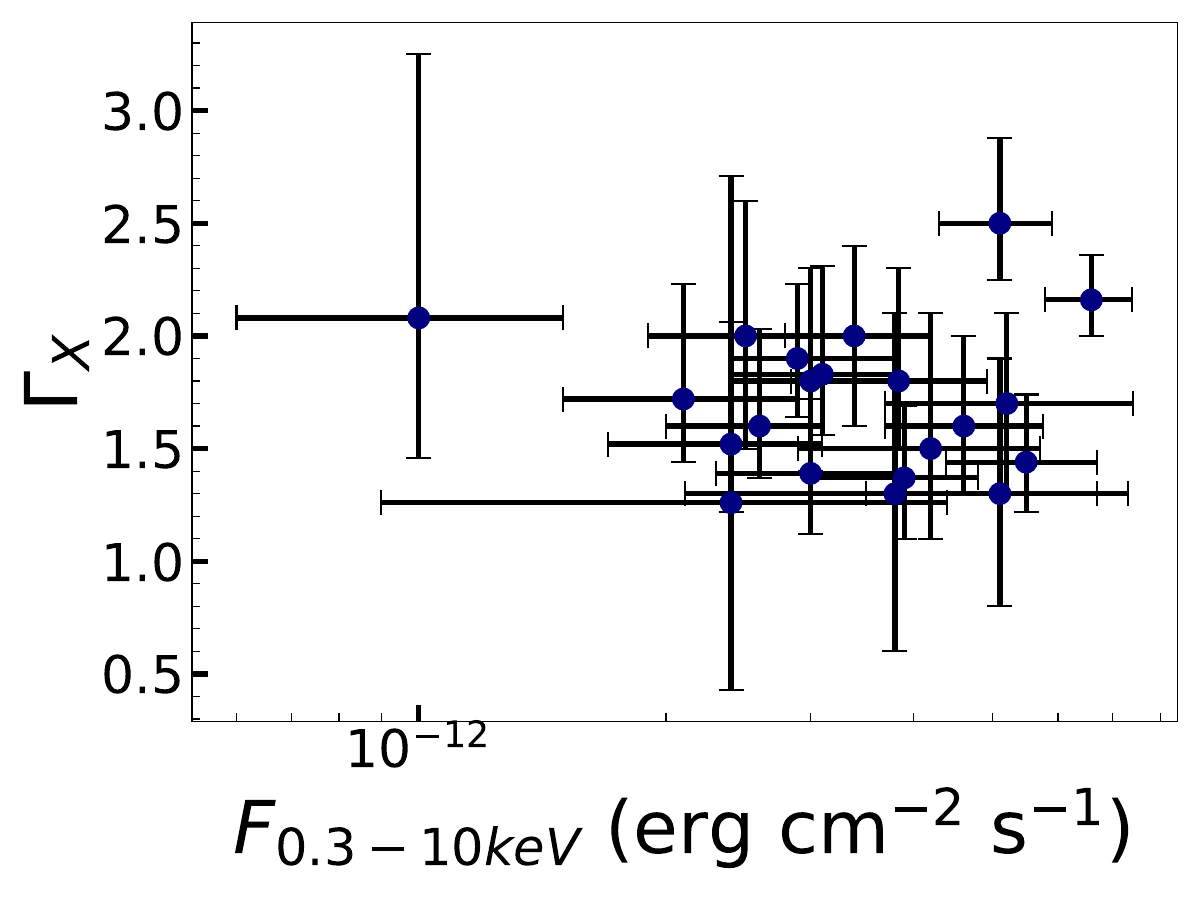}
    \label{fig:4b}}
    \caption{Variation of the observed flux as a function of spectral index. The upper panel shows results from {\it Fermi}-LAT, while the lower panel presents corresponding measurements from {\it Swift}-XRT, enabling a direct comparison of flux–spectral variations across the $\gamma$-ray and X-ray bands.}
    \label{F_ind}
\end{figure}

\begin{figure}
    \centering
    \includegraphics[angle=-90,width=0.99\linewidth]{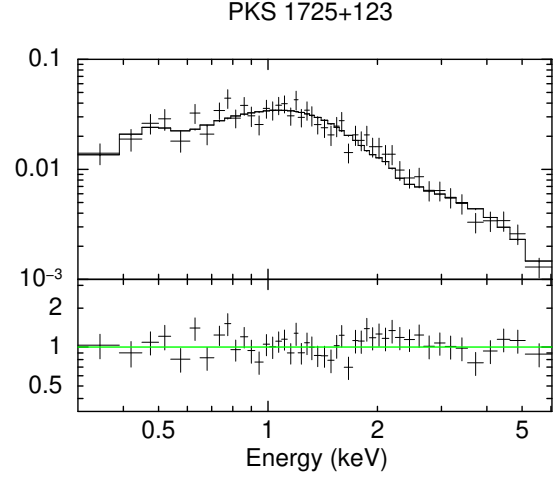}
    \caption{Time averaged {\it Swift}-XRT spectra of PKS 1712+123: The spectra is well fitted by a simple powerlaw model, with, $\Gamma=(1.749\pm0.084)$, column density, $N_H = (9.01\pm2.02)\times 10^{20} cm^{-2}$, $\chi^{2}_{red} = 37.71/44$, and unabsorbed flux = $(3.998\pm0.148)$\ergcm{-12}.}
    \label{XRT_spectra}
\end{figure}



\section{Discussions and CONCLUSIONs}\label{sec:discussions}
AGNs, specifically blazars, are the most numerous detected extragalactic $\gamma$-ray emitters. Their cosmological number density and bolometric luminosity make them ideal sources of cosmic accelerators and UHECRs \citep[see, e.g.,][]{Aloisio:2017qoo}. 
On the other hand, the association with IceCube-detected neutrinos provides smoking-gun evidence for hadronic acceleration, constrained by their multiwavelength SED \citep[e.g.,][]{Murase:2018iyl, Fermi-LAT:2019hte, Giommi:2020hbx, Sahakyan:2022nbz,  Prince:2023bhj}. Separate emission regions for electrons and protons, and a more diffusive proton propagation than the Bohm approximation, can lead to $\sim 100$ PeV neutrinos from blazar jets \citep[see, e.g.,][]{Rodrigues:2025cpm}. Depending on their acceleration and propagation, they may also contribute to the cosmogenic neutrino spectrum \citep[e.g.,][]{Das:2020hev}. The neutrino event IC201021A lies within the 90\% containment region of PKS 1725+123. However, no neutrinos were observed during the flaring state modeled in this work.
We consider a power-law spectrum for electrons and protons with a sharp cutoff, implying that cooling and/or escape dominate above the cutoff energy.

The source was in a low-activity state before March 2025; a rapid flux enhancement was observed by {\it Fermi}-LAT, followed by MWL monitoring by radio to VHE $\gamma$-ray telescopes. The {\it Fermi}-LAT light curve in 2025 shows rapid flux variations, with \texttt{$F_{\rm var}$} and VA values around $\sim$68\% and 50, respectively, while \texttt{$\tau_{\rm var}\leq6$} hours indicates a compact emission region of size $\sim10^{16}$ cm, situated at $\sim10^{17}$ cm from the supermassive black hole. Additionally, these variability timescales can be linked to characteristic timescales, such as the source's cooling timescales. The broadband SED shown in Figure \ref{fig:mar_flare} gives the synchtron peak frequency ($\nu_{\rm syn} = 4\times10^{13}$ Hz), which can be used to estimate the electron Lorentz factor ($\gamma_e$) as given below \citep{1979rpa..book.....R}
\begin{equation}
    \nu_{\rm syn} = 4.2\times10^{6} \cdot \frac{\delta_D\cdot B \cdot \gamma_e^2}{1+z} \,  \text{Hz},
\end{equation}
The synchrotron cooling timescale in the observer's frame in this instance is given as \citep{1979rpa..book.....R}:
\begin{equation}
    t_{\rm syn,cool} \sim 7.7\times10^{8} \cdot \frac{1+z}{B^2 \gamma_e \delta} \, \text{s},
\end{equation}
Similarly, the expression for the IC cooling timescale in the Thomson regime is given as \citep[e.g.,][]{1999MNRAS.306..551C}
\begin{equation}
t_{\mathrm{IC}} = \frac{1+z}{\delta_D} \cdot \frac{3 m_e c}{4 \sigma_T U_{\mathrm{rad}} \gamma_e} \, \text{s}
\end{equation}
where $m_e$ is the mass of electron, $c$ is the speed of light, $\sigma_T$ is the Thomson cross-section and $U_{\mathrm{rad}} = u_{\mathrm{disk}} + u_{\mathrm{BLR}} = 1.75 \ \mathrm{erg \, cm^{-3}}$ is the total radiation density dominated by external photons, obtained in our modeling.

On top of these cooling timescales, we can also estimate the acceleration timescale ($t_{\mathrm{acc}}$) of electrons, assuming the diffusive shock acceleration mechanism \citep[see.,][]{1983RPPh...46..973D, 1987PhR...154....1B, 2000ApJ...536..299K}
\begin{equation}
t_{\mathrm{acc}} \sim 3.79 \times 10^{-7} \, \xi \, \frac{(1+z)\gamma_e}{B \delta} \,s,
\end{equation}
where $\xi$ is the acceleration efficiency parameter. Adopting these values from our broadband SED modelling $B$ = 2 G, $z$ = 0.586, and $\delta_D=30$, we estimate the value of $\gamma_e\sim500$, $t_{\rm syn, cool}\sim5.5$ hours (comparable to the $\tau_{\rm var}$) and $t_{\mathrm{IC}}\sim30$ mins. Assuming the diffusive shock acceleration process is highly efficient ($\xi \sim10^3 - 10^5$), and thus the $t_{\mathrm{acc}}<<1 s$, which is much faster than the $t_{\mathrm{acc}}$ timescales discussed. 
The hierarchy ($\tau_{\rm var}\sim t_{\rm syn, cool} \gg t_{\mathrm{IC}} \gg t_{\mathrm{acc}}$) indicates a rapid cooling phase during the outburst. 
IC losses are energetically dominant, but the observed variability is more closely associated with synchrotron cooling and/or the light-crossing time. 

The $\gamma$-ray flux vs. spectral index (Pearson coefficient:$r = -0.48$, p-value $\sim10^{-3}$, see Figure \ref{fig:4a}) demonstrates a significantly harder-when-brighter pattern, suggesting that $\gamma$-ray variability stems from heightened particle acceleration or injection, resulting in a harder population of high-energy electrons (e.g., inverse Comptonization). Conversely, X-ray flux vs. $\Gamma_X$ as shown in Figure \ref{fig:4b}, shows no trend, suggesting a consistent emission mechanism, possibly driven by a steady particle population or an alternate emission region. Additionally, have performed the X-ray spectral analysis using {\it Swift}-XRT observations in the 0.3-10 keV range. The time-averaged X-ray spectra, shown in Figure \ref{XRT_spectra}, exhibit a stable hard X-ray spectral feature throughout the 2025 avalanche.

An alternative injection spectrum with an exponential cutoff, broken power-law, or log-parabola function is also assumed in the literature.
In our modelling, two external photon fields are required to reproduce the high-energy $\gamma$-ray flux, which we attribute to external IC scattering of accretion-disk and BLR photons. The high-energy spectrum is entirely explained by leptonic IC emission, while the X-ray emission arises from electron SSC. The latter strongly constrains the hadronic contribution. The temperatures of these blackbody fields are adjusted to fit the data. They are higher than that of a standard Shakura–Sunyaev accretion disk in a supermassive black hole \citep{Liu:2022cph}, whose thermal emission typically peaks in the optical/UV. However, there is a hot X-ray corona above and below the disk, whose luminosity is less than, but comparable to, the disk luminosity. The coronal X-rays are also a target for $\gamma\gamma$ pair production \citep{2009MNRAS.397..985G}.  We neglect $\gamma\gamma$ absorption of the IC component on accretion-disk photons, since in our model the $\gamma$-ray spectrum extends to $\lesssim10$ GeV. 

H.E.S.S. and MAGIC have detected multi-TeV $\gamma$-rays during the flare. Interpreting this VHE emission as leptonic emission from the source is difficult due to strong EBL absorption at such high energies. The detection of a steady TeV flux would instead favor a cosmogenic origin due to the interaction of the highest-energy cosmic rays with the cosmic microwave background and EBL during extragalactic propagation \citep{2010APh....33...81E, Razzaque_2012, Das:2019gtu} and can also constrain synchrotron pair echo emission \citep[see, e.g.,][]{Murase:2011yw}. We show the sensitivity of the northern array of Cherenkov Telescope Array \citep[CTAO-N;][]{Gueta:2021vO} and Large High-Altitude Air Shower Observatory \citep[LHAASO;][]{Vernetto:2016gro} in Fig.~\ref{fig:mar_flare}. Future TeV $\gamma$-ray observations with CTA and LHAASO will further constrain cosmic-ray production in this source.

The X-ray data constrain $\gamma^\prime_{p,\rm max}$ in our model, beyond which escape can dominate for a rigidity-dependent diffusion (See Fig.~\ref{fig:qe_sources}. At higher proton energies beyond the cutoff, photopion interactions with BLR photons become efficient, producing $\pi^0$-decay $\gamma$-rays that trigger electromagnetic cascades and secondary $e^\pm$ production, leading to an X-ray excess that further limits the allowed neutrino flux. However, the Bethe-Heitler interaction occurs efficiently with both photon fields. For the duty cycle of the flaring period during MJD 60748 - 60955, we estimate a neutrino event rate of $\approx0.3$ yr$^{-1}$. Simultaneous observation during the quiescent state may tighten this limit, since the hadronic contribution and hence neutrino flux level could be lower. In addition, it should be noted that there may be degeneracy in model parameters, as well as uncertainty in the size and location of the emission region. However, our treatment of two external photon fields explains the neutrinos originating from $p\gamma$ interactions in the hot accretion disk/corona region, while BLR photons primarily serve as targets for IC emission by electrons. Our one-zone leptohadronic modelling of the 2025 flares suggests that the $\gamma$-ray emission is primarily driven by leptonic radiation processes, with hadronic (electromagnetic cascade, as shown by the orange line in Fig.~\ref{fig:mar_flare}) interactions being less significant. Although the cascade flux is aligned with the archival quiescent $\gamma$-ray data (in gray), the lack of concurrent X-ray observation during the quiescent phase complicates the origin of this emission. 




\bibliographystyle{mnras}
\bibliography{Refs}


\appendix

\section{Numerical analysis of hadronic cascade in the comoving jet frame}\label{app:hadronic}

\begin{figure*}
    \centering
    \includegraphics[width=0.33\textwidth]{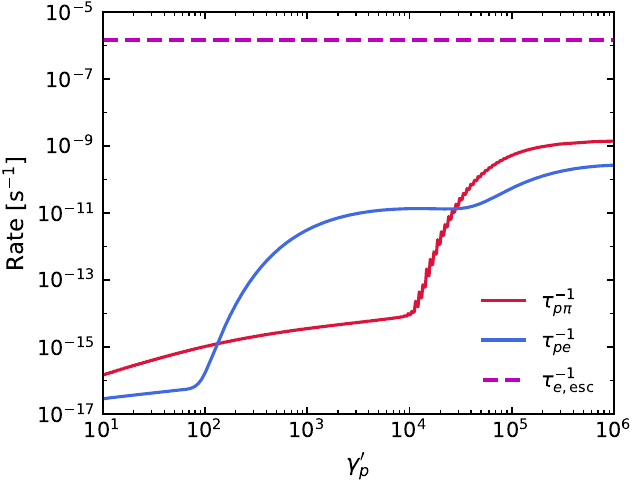}%
    \includegraphics[width=0.33\textwidth]{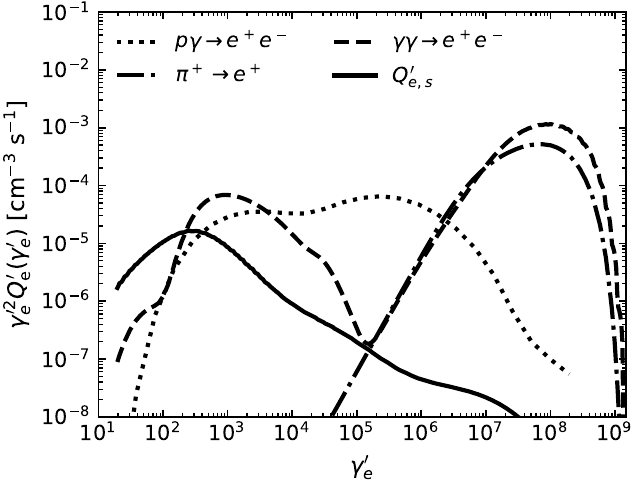}%
    \includegraphics[width=0.33\textwidth]{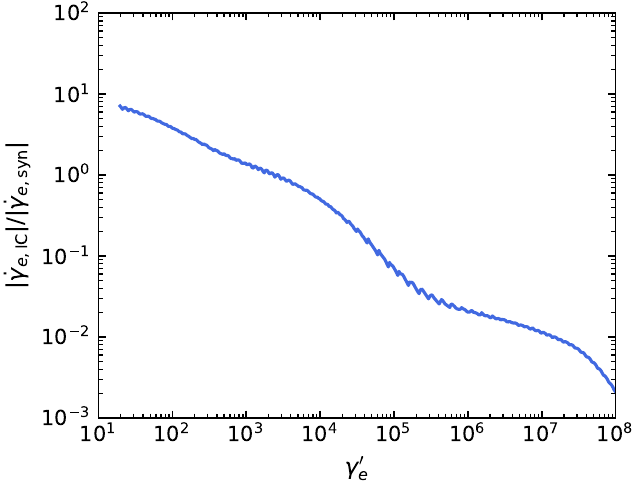}
    \caption{\textit{Left:} Photohadronic interaction rates and the electron escape rate in the comoving jet frame for the leptohadronic model of PKS 1725+123. It is computed by integrating over the photon distribution, using the energy-dependent cross section and inelasticity in the proton rest frame \citep{Stecker:1968uc, Chodorowski_92, Mucke_00}. \textit{Middle:} Spectrum of secondary electrons injected by hadronic interactions and the steady-state secondary electron spectrum for the proton spectrum $dN/d\gamma^\prime_p\sim \gamma^{\prime -2}_p$ in the range $\gamma^\prime_p=10-10^6$. \textit{Right:} IC to synchrotron loss rate of the steady state secondary electron spectrum using the total target photon field in the steady state (see Eqn.~\ref{eqn:target}).}
    \label{fig:qe_sources}
\end{figure*}

The differential number densities per unit energy of accretion disk and BLR photons are denoted by $n_{\rm BLR}(\epsilon^\prime)$ and $n_{\rm disk}(\epsilon^\prime)$, where $\epsilon^\prime = h\nu^\prime/m_ec^2$ is the dimensionless photon energy. In contrast to the soft-photon field produced by the primary electrons, which is orders of magnitude weaker, these external fields, Doppler-boosted in the jet frame, provide the dominant target photons for $p\gamma$ interactions. 
The left panel of Fig.~\ref{fig:qe_sources} shows the interaction rates as functions of the proton Lorentz factor $\gamma^\prime_p$ in the comoving frame. We denote the photopion production timescale by $\tau_{p\pi}$ and the Bethe--Heitler (BH) pair-production timescale by $\tau_{\rm pe}$. For comparison, we also show the electron escape timescale, taken to be comparable to the dynamical timescale, $\tau_{e,\rm esc} = 2R^\prime/c$. The proton escape timescale can be much longer because of diffusion, and efficient photohadronic interactions in our model require $\tau_{p,\rm esc} \sim 10^3 R^\prime/c$ at $\gamma^\prime_p \approx 10^6$. More generally, an energy-dependent parametrization may be written as $t_{\rm esc} \gtrsim 10^3 (R^\prime/c)(E^\prime_p/1,\text{PeV})^{-1}$ over the proton energy range relevant for interactions inside the jet. We assume that above this range, either the acceleration efficiency declines or protons escape more efficiently because of rigidity-dependent diffusion. The spectra of secondary particles are weighted by the corresponding interaction rates. For example, the $\pi^0$-decay $\gamma$-ray spectrum is weighted by $R_{p\pi}/R_{\rm tot}$, where $R_{p\pi}$ is the photopion interaction rate and $R_{\rm tot} = R_{\rm BH} + R_{p\pi}$ is the total interaction rate.

Relativistic electrons from BH pair production $Q'_{e,\rm BH}$, decay of charged-pion $Q^\prime_{e,\pi}$, and $\gamma\gamma$ interactions $Q^\prime_{e,\gamma\gamma}$ are injected into the secondary hadronic cascade. $Q'_{e,\rm BH}$, $Q^\prime_{e,\pi}$ and $\pi^0$ decay $\gamma$-ray spectrum $Q^\prime_{\gamma,\pi}$ are calculated using the formalism in \cite{Kelner:2008ke}. High-energy $\gamma$ rays are attenuated through $\gamma\gamma \rightarrow e^\pm$ interactions with soft photons in the external radiation field. The escaping TeV spectrum is therefore suppressed according to $Q^\prime_{\gamma,\rm esc}(\epsilon^\prime)=Q^\prime_{\gamma,\pi}(\epsilon^\prime),(1-e^{-\tau_{\gamma\gamma}})/\tau_{\gamma\gamma}$, where $\tau_{\gamma\gamma}$ is the optical depth, computed following \citet{Gould_1967} by integrating the relevant cross section over the target photon distribution. The injection spectrum of pairs generated through $\gamma\gamma$ absorption, $Q^\prime_{e,\gamma\gamma}$, is calculated numerically using the formalism of \citep{Boettcher_1997}. The various source terms are shown in the middle panel of Fig.~\ref{fig:qe_sources} for the proton spectrum assumed in this article. The evolution of the isotropic electron distribution $N^\prime_e(\gamma_e^\prime)$ is governed by the stationary one-dimensional Fokker-Planck equation,
\begin{align}
\frac{\partial}{\partial\gamma_e^\prime}&\left[(\dot{\gamma}_{e,\rm syn}(\gamma_e^\prime)
    + \dot{\gamma}_{e,\rm IC}(\gamma_e^\prime))N^\prime_{e,s}(\gamma_e^\prime)\right]  =  \nonumber \\
    &Q^\prime_{\rm BH}(\gamma_e^\prime) + Q^\prime_{\pi^+}(\gamma_e^\prime) + Q^\prime_{\gamma\gamma}(\gamma_e^\prime)
    - \frac{N^\prime_{e,s}(\gamma_e^\prime)}{\tau_{e,\rm esc}},
    \label{eqn:fokker}
\end{align}
where $\dot{\gamma}_{e,\rm syn}(\gamma_e^\prime)$ and $\dot{\gamma}_{e,\rm IC}(\gamma_e^\prime)$ are the synchrotron and IC cooling rates in the jet frame, respectively, and the source term is given as
\begin{align}
    Q^\prime_{e,\rm inj} =  Q^\prime_{\rm BH}(\gamma_e^\prime) + Q^\prime_{\pi^+}(\gamma_e^\prime) + Q^\prime_{\gamma\gamma}(\gamma_e^\prime) \ .
\end{align}
We recast Eqn.~\ref{eqn:fokker} as a one-dimensional Volterra equation and solve it by downward integration in $\gamma$ to obtain the steady-state secondary electron distribution $N^\prime_{e,s}(\gamma_e^\prime) \simeq Q^\prime_{e,s}\tau_{e, \rm esc}$ as,
\begin{equation}
N^\prime_e(\gamma_{e,s}^\prime)
=
\frac{1}{\Lambda(\gamma_e^\prime)}
\int_{\gamma_e^\prime}^{\infty}
\left[
Q^\prime_{e,\rm inj}(\tilde{\gamma}_e)
-
\frac{N^\prime_{e,s}(\tilde{\gamma}_e)}{\tau_{e,\rm esc}}
\right]
\,d\tilde{\gamma}_e
\label{eqn:volterra}
\end{equation}
where, $\Lambda(\gamma_e^\prime)=\left|\dot{\gamma}_{e,\rm syn}(\gamma_e^\prime)+\dot{\gamma}_{e,\rm IC}(\gamma_e^\prime)\right|$ is the total cooling rate.
We extend the integral solution method of \cite{2013ApJ...768...54B}, which includes synchrotron losses $\dot{\gamma}_{e,\rm syn}(\gamma_e^\prime) = -\nu_0 \gamma_e^{\prime2}$, by incorporating the full IC cooling rate, $\dot{\gamma}_{e,\rm IC}(\gamma_e^\prime)$, from the evolving photon field into the total loss term. Specifically, for a given electron Lorentz factor $\gamma_e^\prime$ and total photon field $n_{\rm ph,tot}(\epsilon^\prime)$, we evaluate
\begin{align}
\dot{\gamma}_{e,\rm IC}(\gamma_e^\prime) = -\frac{c}{m_e c^2}
\int {\rm d}\epsilon_s\, \epsilon_s
\int {\rm d}\epsilon\, n_{\rm ph,tot}(\epsilon)\,
\frac{{\rm d}\sigma_{\rm IC}}{{\rm d}\epsilon_s}(\epsilon,\epsilon_s,\gamma_e^\prime),
\end{align}
using the Klein--Nishina differential cross section $d\sigma_{\rm IC}/d\epsilon$ \citep{Blumenthal_1970}. In a first pass, we ignore IC cooling and solve the steady-state Eqn.~\ref{eqn:volterra} using only synchrotron losses, obtaining an initial approximation $N_{e,s}^{\prime(0)}(\gamma_e^\prime)$. From this, we compute the cascade synchrotron emissivity \citep{2013ApJ...768...54B}
\begin{align}
Q_{\rm syn}^{\prime(0)}(\epsilon^\prime) = A_0 \epsilon'^{-3/2} \int_1^\infty N^{(0)}_{e,s}(\gamma'_e) \gamma'^{-2/3}_e e^{-\epsilon^\prime / b\gamma'^2_e} \, d\gamma'_e
\end{align}
with 
$A_0 =c\sigma_TB'^2/[6\pi m_e c^2 \Gamma(4/3)b^{4/3}]$ being a normalization constant, 
$b=B'/B_{\rm crit}$ and $B_{\rm crit} = 4.4\times 10^{13}$ G. The corresponding photon density per unit energy is $n_{\rm syn,cas}^{(0)}(\epsilon^\prime)$, and we construct the total target photon field
\begin{equation}
n_{\rm tot}^{(0)}(\epsilon^\prime) = n_{\rm BLR}(\epsilon^\prime)
+ n_{\rm disc}(\epsilon^\prime)
+ n_{\rm syn, prim}(\epsilon^\prime)
+ n_{\rm syn, cas}^{(0)}(\epsilon^\prime).
\label{eqn:target}
\end{equation}
where $n_{\rm syn, prim}$ is the synchrotron radiation of primary electrons injected in the comoving jet frame. We use the primary synchrotron spectrum from our epoch-1 as a representative case in the middle panel of Fig.~\ref{fig:qe_sources}.
Using $n_{\rm ph,tot}^{(0)}$, we calculate $\dot{\gamma}_{e,\rm IC}^{(0)}(\gamma_e^\prime)$ and update the loss coefficient $\Lambda^{(0)}(\gamma_e^\prime) = \nu_0 \gamma_e^{\prime2} + |\dot{\gamma}_{e,\rm IC}^{(0)}(\gamma_e^\prime)|$. In the second pass, we re-solve the steady-state equation with this full loss coefficient, yielding an improved electron distribution $N_{e,s}^{(1)}(\gamma_e^\prime)$. This gives a new synchrotron photon field $n_{\rm syn, cas}^{(1)}(\epsilon^\prime)$, and hence, a new total photon field $n_{\rm ph,tot}^{(1)}(\epsilon^\prime)$, and thus a new IC cooling rate $\dot{\gamma}_{e,\rm IC}^{(1)}$.
We repeat this procedure for several iterations,
\begin{equation}
N_{e,s}^{\prime(k)}(\gamma_e^\prime) \;\rightarrow\; n_{\rm ph,tot}^{(k)}(\epsilon^\prime)
\;\rightarrow\; \dot{\gamma}_{e,\rm IC}^{(k)}(\gamma_e^\prime)
\;\rightarrow\; N_{e,s}^{\prime(k+1)}(\gamma_e^\prime),
\end{equation}
until the electron distribution converges, i.e.\ the relative change
$|N_{e,s}^{\prime(k+1)}(\gamma_e^\prime) - N_{e,s}^{\prime(k)}(\gamma_e^\prime)|/N_{e,s}^{\prime(k)}(\gamma_e^\prime)$ falls below a prescribed tolerance over the $\gamma_e^\prime$ range that dominates the emission. The final converged distribution, $N^\prime_{e,s}(\gamma_e^\prime)$, is then used to compute the steady-state cascade synchrotron emissivity and the IC emissivity, $Q^\prime_{\rm IC}(\epsilon^\prime)$, from the full photon field. These together constitute the hadronic cascade component of our spectral energy distribution model, superposed on the radiation from the steady-state primary electron background. The right panel of
Fig.~\ref{fig:qe_sources} shows the ratio of the inverse-Compton to synchrotron cooling rates,
$|\dot{\gamma}_{e,\rm IC}|/|\dot{\gamma}_{e,\rm syn}|$, as a function of the electron
Lorentz factor $\gamma_e^\prime$.
At low $\gamma_e^\prime$, the ratio is of order unity or larger, indicating that IC and synchrotron losses are comparable or IC-dominated.
As $\gamma_e^\prime$ increases, the ratio steadily decreases by several orders of magnitude,
so that, at the highest Lorentz factors, electrons lose energy predominantly through synchrotron radiation.

\bsp	
\label{lastpage}
\end{document}